\documentclass[final, aps, reprint, preprintnumbers, footinbib, showpacs, citeautoscript, superscriptaddress, balancelastpage]{revtex4-1}

\preprint{In Preparation}

\usepackage[T1]{fontenc}
\usepackage[english]{babel}

\usepackage{amsmath}
\usepackage{amssymb}
\usepackage{braket}

\usepackage{textcomp}
\usepackage{times}
\usepackage{txfonts}

\usepackage{graphicx}
\usepackage[dvipsnames]{xcolor}

\usepackage[colorlinks, urlcolor=blue, citecolor=blue, linkcolor=blue, pdfstartview=FitH]{hyperref}
\usepackage[all]{hypcap}

\usepackage[normalem]{ulem} 

\begin{document}

\def\affiEII{Experimentelle\ Physik\ 2, Technische\ Universit\"at\ Dortmund, D-44221 Dortmund, Germany}
\def\affiEIII{Experimentelle\ Physik\ 3, Technische\ Universit\"at\ Dortmund, D-44221 Dortmund, Germany}
\def\affiSOLAB{Spin\ Optics\ Laboratory, Saint~Petersburg\ State\ University, 198504 St.~Peterbsurg, Russia}
\def\affiIOFFE{Ioffe\ Institute, Russian\ Academy\ of\ Sciences, 194021 St.~Petersburg, Russia}

\title{Nuclear spin cooling by helicity-alternated optical pumping at weak magnetic fields in \emph{n}-GaAs}

\author{P.~S.~Sokolov}
\affiliation{\affiEII}
\affiliation{\affiSOLAB}
\author{M.~Yu.~Petrov}
\affiliation{\affiSOLAB}
\author{K.~V.~Kavokin}
\affiliation{\affiSOLAB}
\affiliation{\affiIOFFE}
\author{A.~S.~Kurdyubov}
\author{M.~S.~Kuznetsova}
\author{R.~V.~Cherbunin}
\author{S.~Yu.~Verbin}
\affiliation{\affiSOLAB}
\author{N.~K.~Poletaev}
\affiliation{\affiIOFFE}
\author{D.~R.~Yakovlev}
\affiliation{\affiEII}
\affiliation{\affiIOFFE}
\author{D.~Suter}
\affiliation{\affiEIII}
\author{M.~Bayer}
\affiliation{\affiEII}
\affiliation{\affiIOFFE}
\date{\today}

\begin{abstract}
The spin dynamics of localized donor-bound electrons interacting with the nuclear spin ensemble in $n$-doped GaAs epilayers is studied using nuclear spin polarization by light with modulated circular polarization.
We show that the observed build-up of the nuclear spin polarization is a result of competition between nuclear spin cooling and nuclear spin warm-up in the oscillating Knight field. 
The developed model allows us to explain the dependence of nuclear spin polarization on the modulation frequency and to estimate the equilibration time of the nuclear spin system that appears to be shorter than the transverse relaxation time $T_2$ determined from nuclear magnetic resonance. 
\end{abstract}


\maketitle

\section{\label{sec:intro}Introduction}

In semiconductors in which lattice nuclei have non-zero spins, the electron-nuclear hyperfine interaction limits the electron spin coherence unless the nuclear spin system is polarized up to a high degree. 
On the other hand, dynamically polarized nuclear spins can create a strong effective magnetic field, namely the Overhauser field, $B_N$, acting upon electron spins. 
Therefore, control of the nuclear spin polarization by, e.g., time-shaped optical or electric pumping may have application potential~\cite{HennebergerQBits}, and the time scales on which the Overhauser field develops and changes deserve thorough investigation~\cite{DyakonovSpinBook, UrbaszekRMP13}.

Since the nuclear spin system of a solid is relatively weakly coupled to the crystal lattice, in many cases it reaches a thermal equilibrium state characterized by a spin temperature~\cite{GoldmannSpinTemp} that is different from the lattice temperature~\cite{OO}. 
The internal equilibrium in the nuclear spin system is established via magnetic dipole-dipole interactions on the time scale of $T_2 \sim 10^{-4}$~s~\cite{PagetPRB77}. 
The spin temperature can be many orders of magnitude lower, by absolute value, than the lattice temperature~\cite{MerkulovFTT98}, and it can be both positive or negative~\cite{KalevichIzv82, VladimirovaArXiv17}. 
Its evolution in the absence of pumping is governed by various processes of spin-lattice relaxation, for example, via a quadrupole mechanism~\cite{PagetPRB08,EickhoffPRB02,KoturPRB16} and, in the case of spatial inhomogeneity, affected also by spin diffusion~\cite{GiriPRL13, MalinowskiPRL17}.

The dynamics of the nuclear spin temperature under optical pumping is even more complex, because in this case nuclear spins interact with non-equilibrium electrons. 
It is known, that the efficiency of nuclear spin cooling by circularly polarized light decreases when the degree of circular polarization is modulated. 
Obviously, nuclear spins can be efficiently cooled if the modulation frequency is smaller than $T_1^{-1}$. 
However, cooling is possible also at higher frequencies, $\omega \gg T_1^{-1}$, because of the Knight field created by photoexcited electrons, which alternates at the same frequency as the electron mean spin \cite{OO}. 
In the case of high-frequency polarization modulation, i.\,e., at frequencies much larger than $T_2^{-1}$, no cooling is possible unless the modulation frequency is close to the Nuclear Magnetic Resonance (NMR) frequency in the applied magnetic field. 
In this latter case, nuclear spins can be pumped via so-called resonant cooling~\cite{OO}.
At even higher frequencies of intensity modulation, implemented by pulsed lasers with the pulse repetition rate exceeding 75~MHz, the resonantly driven electron spin system can be prepared in a highly excited state maintained at a large transverse magnetic field.
In this case, the phase relaxation of the electron spin precession might be treated as a change of the temperature of the electron spin subsystem, and, to equilibrate the temperature balance, the nuclear spin system is cooled inducing a considerably large Overhauser field~\cite{KikkawaSci00}.
Moreover, for strongly localized electron spins, e.g., in the ensemble of singly charged quantum dots, in the regime of electron-spin mode locking~\cite{GreilichSci06} the Overhauser field provides a channel for the frequency focusing of the electron spin coherence~\cite{GreilichSci07} considered as a method of decoupling the electron spin from the nuclear spin ensemble that is an alternative to deep cooling.

While polarization of a single spin can reach 99\%~\cite{FalkPRL15}, the optimal strategy to achieve a highly polarized mesoscopic nuclear spin state is currently a subject of discussion while the current record in a single quantum dot is 80\%~\cite{ChekhovichNmat17}, and at least an order of magnitude lower in quantum dot ensembles or bulk semiconductors.
The efficiency of nuclear spin optical cooling in the intermediate frequency range $T_1^{-1} < \omega < T_2^{-1} $ has not been investigated either experimentally or by a quantitative theory. 
This paper aims to fill this gap.

Toward that end, a method originally developed to investigate ``spin inertia''~\cite{HeisterkampPRB15} is adapted to examine the nuclear spin dynamics in $n$-doped GaAs. 
As we will show, the Knight field oscillating synchronously with the electron mean spin indeed provides an off-resonant cooling of the nuclear spin system up to frequencies of the order of $T_2^{-1}$. 
On the other hand, the oscillating Knight field warms up the nuclear spin system. 
The competition of these two processes results in a cut-off frequency of nuclear spin cooling $\omega_{1/2}$, lying in between $T_1^{-1}$ and $T_2^{-1}$. 
Knowing the parameters of the electron spin system, one can use the measured $\omega_{1/2}$ to determine the parameters of the nuclear spin correlator, primarily the value of $T_2$ in weak magnetic fields, which cannot be directly measured using standard NMR techniques.

\section{\label{sec:experiment}Experimental Results}

The studied sample is an $n$-doped GaAs epitaxial layer grown by liquid-phase epitaxy on top of a semi-insulating (001) GaAs substrate. 
The $20$~\textmu{}m epitaxial layer was doped by Si providing a donor concentration $n_\mathrm{d} = 4\times10^{15}$~cm$^{-3}$~\cite{DzhioevPRL02, DzhioevPRB02}. 
All measurements are done at the sample temperature $T = 1.6$~K.
The photoluminescence (PL) is excited by a tunable Ti:Sapphire laser operating at $E_\mathrm{exc} = 1.540$~eV corresponding to the absorption edge of the GaAs band-to-band transition. 
The laser is focused on the sample surface through an achromatic doublet (focal distance $F=200$~mm) into a spot of about $80$~\textmu{}m in diameter ($1/e^2$ width) and the PL is collimated with the same lens throughout all subsequent measurements.
The helicity of optical excitation is controlled by an electro-optical modulator driven by a radio-frequency harmonic oscillator that is used to avoid a possible impact of higher-frequency harmonics.
The time-dependent phase shift of the optical frequency is converted into a linear polarization modulation, which is further transformed into a modulation of the circular polarization degree by a following quarter-wave plate.
This allows us to implement excitation protocols with a fast continuous switch between circular right ($\sigma^+$) and circular left ($\sigma^-$) light polarizations. 
The PL is collected in reflection geometry, spectrally filtered by an $0.125$~m fixed-slit monochromator eliminating the residual scattered light, and dispersed by a $0.5$~m single-grating spectrometer followed by a gated single-photon counter. 
The analysis of the circular polarization degree of the PL is done by a photo-elastic modulator followed by a Glan-Taylor polarizer. 
The intensities of the circular left and circular right PL polarization components are detected with a two-channel photon counting device.
The degree of circular polarization is obtained as $\rho_c = (I^\mathrm{co}-I^\mathrm{cross})/(I^\mathrm{co}+I^\mathrm{cross})$ with the intensities $I^\mathrm{co}$ and $I^\mathrm{cross}$ detected at co-circular and cross-circular PL polarization helicities with respect to the excitation. 
The accurate gating of the $I^\mathrm{co}$ and $I^\mathrm{cross}$ intensities is provided by precise time protocols operated using digital delay electronics synchronized to the gating of the polarization detection scheme such that $\rho_c$ is accumulated only when the system is illuminated with light reaching a circular polarization degree above 80\% during a single half-period of modulation.
In some experiments, to eliminate the possible impact of the nuclear spin polarization, the helicity of the pumping light $(\sigma^+/\sigma^-)$ is modulated at a high frequency $f_\mathrm{mod}$ exceeding several tens of kHz.

\begin{figure}[t]
\includegraphics[clip, width=\columnwidth]{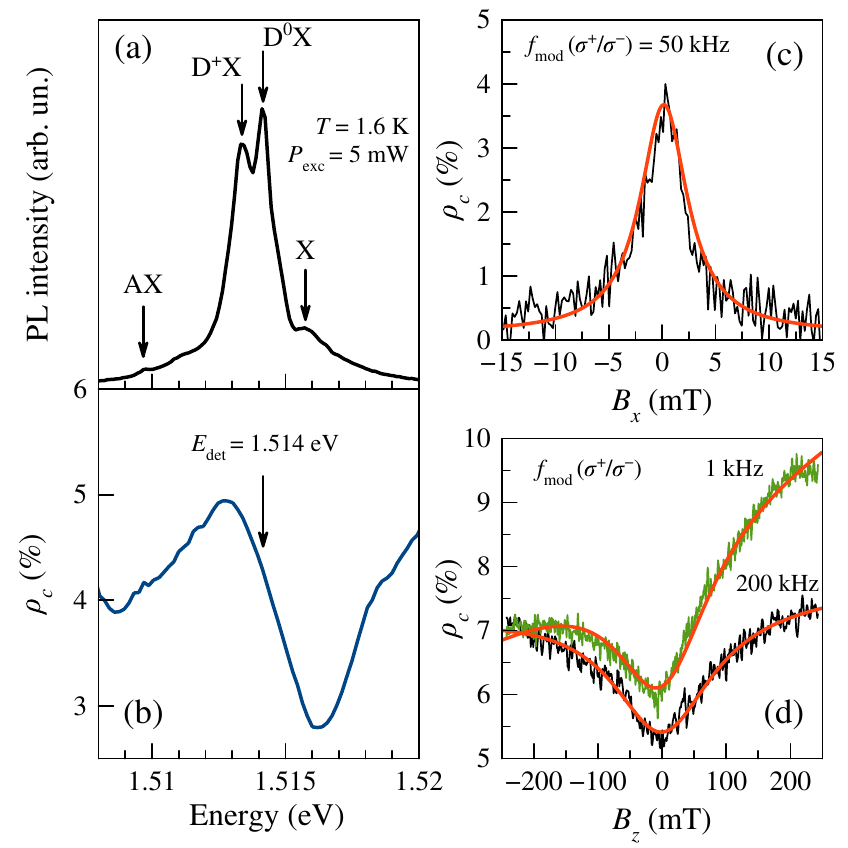}
\caption{(Color online) (a)\ PL spectrum (excitation energy $E_\mathrm{exc}=1.55$~eV) of $n$-doped GaAs measured at $B=0$~T.
(b)\ Spectral dependence of the PL circular polarization degree.
(c)\ Circular polarization degree versus transverse magnetic field (Voigt geometry, $B_z = 0$) measured at alternating helicity of excitation $f_\mathrm{mod} = 50$~kHz. The electron spin relaxation time $\tau_s = 20$~ns is evaluated.
(d)\ Magnetic field dependencies of the circular polarization degree (Faraday configuration, $B_x = 0$) measured for the D$^0$X transition ($E_\mathrm{det} = 1.514$~eV) at fast modulation of the helicity of excitation $f_\mathrm{mod} = 200$~kHz (black curve) and $f_\mathrm{mod} = 1$~kHz (green curve). The corresponding electron correlation time $\tau_c = 310$~ps. Solid lines in panels (c) and (d) result from fitting with Eqs.~\eqref{eq_FastHanle} and \eqref{eq:av_electron}, respectively.
}\label{fig:One}
\end{figure}

The PL spectrum shown in Fig.\ \ref{fig:One}(a) has four distinct peaks corresponding to a recombination of the exciton (X), the exciton bound on neutral and charged donors (D$^0$X and D$^+$X), as well as the exciton acceptor complex (AX). 
The spectrum also demonstrates a non-monotonic behavior of the PL circular polarization degree, as shown in Fig.\ \ref{fig:One}(b). 
The spin polarization is governed mostly by electrons localized on donors \cite{OO}, and the polarization time $T_{1e}$ rapidly increases with increasing distance of the nucleus from the donor center. 
Following Refs.\ \onlinecite{OO} and \onlinecite{KoturPRB16}, the part of the spectrum corresponding to the D$^0$X transition at $E_\mathrm{det} = 1.514$~eV is further analyzed in the magnetic field. 
The choice of this spectral energy is motivated by minimizing the field-independent offset of the PL polarization and obtaining a maximal deviation of $\rho_c$ from its equilibrium value detected at zero field, $\rho_0$, when the magnetic field is applied.

A magnetic field applied along the light propagation axis (Faraday geometry) increases $\rho_c$, an effect known as polarization recovery (PR) [Fig.\ \ref{fig:One}(d)]. 
On the contrary, the application of a transverse magnetic field (Voigt geometry) leads to a decrease of $\rho_c$ with increasing field due to the Hanle effect [Fig.\ \ref{fig:One}(c)]. 
Such a behavior is typical for $n$-doped GaAs and allows one to determine the characteristic values of the electron correlation time, $\tau_c$, and the electron spin relaxation time, $\tau_s$.

Since at high enough modulation frequency $f_\mathrm{mod} \gg T_2^{-1}$ the nuclear spin polarization is negligible \cite{DyakonovSpinBook}, the Hanle curve shown in Fig.\ \ref{fig:One}(c) is, to a good approximation, a Lorentzian:
\begin{equation}
	\rho_c (B_x) = \frac{\rho_0}{1 + B_x^2/B_{1/2}^2}
	\label{eq_FastHanle}
\end{equation}
with $\rho_0 = 0.036$ and $B_{1/2} = 2.7$~mT.
The corresponding spin relaxation time for steady state conditions is evaluated as $(\tau^\ast)^{-1} = \tau^{-1} + \tau_s^{-1}$ where $\tau^\ast =\hbar / \mu_B \vert g_e \vert  B_{1/2}$, $\mu_B=9.274\times10^{-24}$~JT$^{-1}$ is the Bohr magneton, $\vert g_e\vert=0.44$ \cite{WeisbuchPRB77} is the electron $g$~factor. 
Thus, we get the electron lifetime $\tau =10$~ns~\cite{PagetPRB77} taking into account that $\tau_s = 20$~ns for our conditions as evaluated from the fitting of the experimental data in Fig. \ref{fig:One}(c). 

To evaluate $\tau_c$, the dependence of the $\rho_c$ in Fig.~\ref{fig:One}(d) on the longitudinal magnetic field $B_z$ is measured.
In this case, the electron spin $z$~component increases due to a change of the spin relaxation time~\cite{WeisbuchPRB77}.
Supposing a simple field dependence~\cite{DyakonovSpinBook}: $\tau_s' = \tau_s (1 + \mu_B^2 g_e^2 B_z^2 \tau_c^2/\hbar^2)$ and $\rho_c = \rho_\infty \tau_s'/(\tau + \tau_s')$, we obtain
\begin{equation}
\rho_c (B_z) =  \frac{\rho_\infty}{1+\frac{\tau}{\tau_s} \left[ 1+ (\mu_B g_e B_z \tau_c/\hbar)^2 \right]^{-1}}
\label{eq:av_electron}	
\end{equation}
where $\rho_\infty$ is the polarization degree reached in the limit of large magnetic fields.

We have traced the dependence of $\rho_c$ on the longitudinal magnetic field up to $B_z = 0.25$~T [see  Fig.\ \ref{fig:One}(d)].
The electron spin polarization saturates at a certain value providing a PR dependence that represents a wide inverted Lorentzian curve given by Eq.~\eqref{eq:av_electron}. The extracted correlation time of the donor bounded electron is $\tau_c=310$~ps. 
A small additional linear asymmetry of the saturating PR amplitude at high enough positive and negative fields due to equilibrium paramagnetic polarization of electron spins is also observed, however a fast modulation of the pump helicity at $f_\mathrm{mod} = 200$~kHz removes this effect, as shown in Fig.~\ref{fig:One}(d).
The width of the obtained PR curve does not depend strongly on the frequency $f_\mathrm{mod}$ in the range of $0.9$--$12$~kHz, allowing us to conclude that the condition of short correlation time $\tau_c \ll \tau_s$~\cite{DzhioevPRL02} is fulfilled throughout the experiments reported in the rest of this work.

\begin{figure}[t]
\includegraphics[clip, width=\columnwidth]{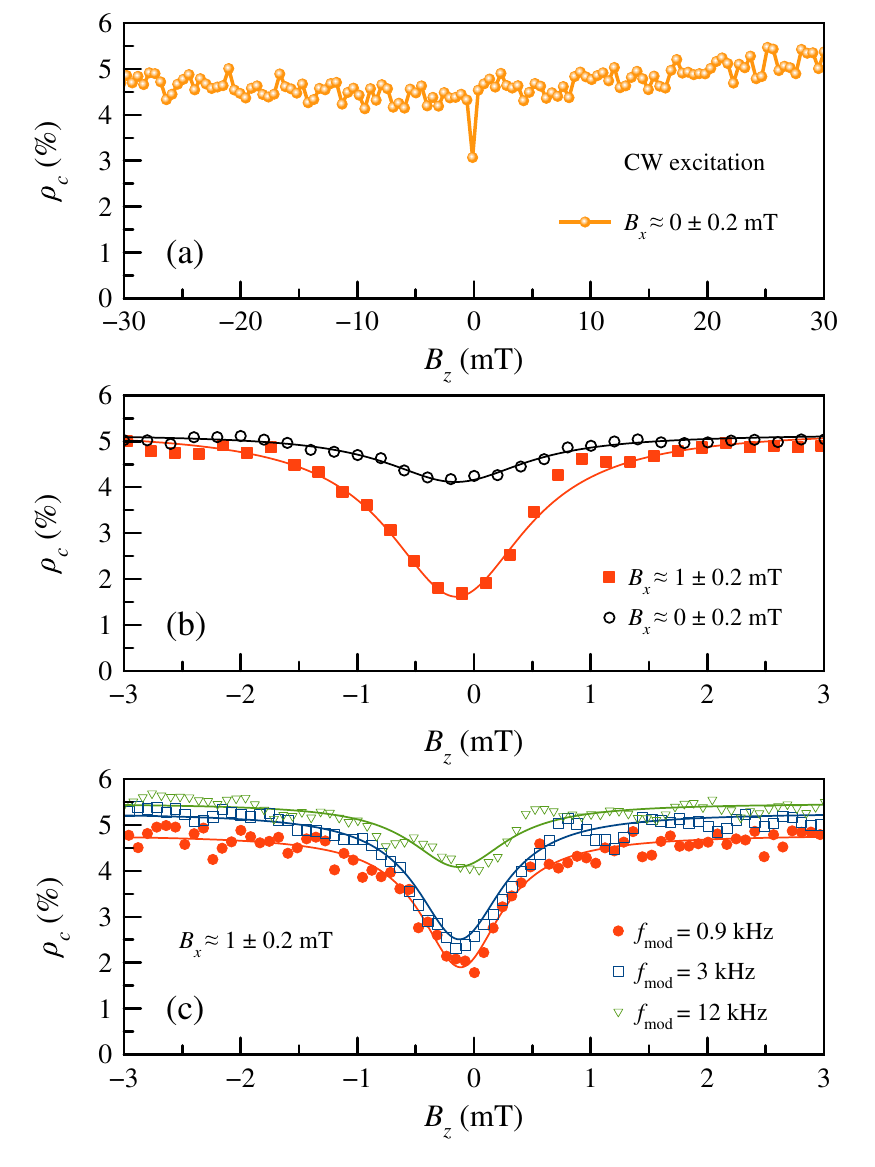}
\caption{(Color online) (a) Manifestation of the additional tiny PR signal in small longitudinal fields (Faraday configuration, $B_x = 0$) measured for the D$^0$X transition ($E_\mathrm{det} = 1.514$~eV) at continuous-wave (CW) excitation. (b) Recovery of the PL circular polarization degree by a longitudinal magnetic field in absence of a transverse magnetic field (circles) and in the presence of $B_x = 1$~mT (squares). (c) PR curves measured at different frequencies of modulation in the presence of a transverse magnetic field. Solid lines in panels (b) and (c) result from fitting with a bell-shaped function: $\rho_c (B_z)=  \rho_\mathrm{sat}-A_\mathrm{PR}/ \left( 1 + B_z^2 / \Delta_B^2 \right)$, with $\rho_\mathrm{sat}$, $A_\mathrm{PR}$, and $\Delta_B$ being fitting parameters. 
}\label{fig:Two}
\end{figure}

\begin{figure}[t]
\includegraphics[clip, width=\columnwidth]{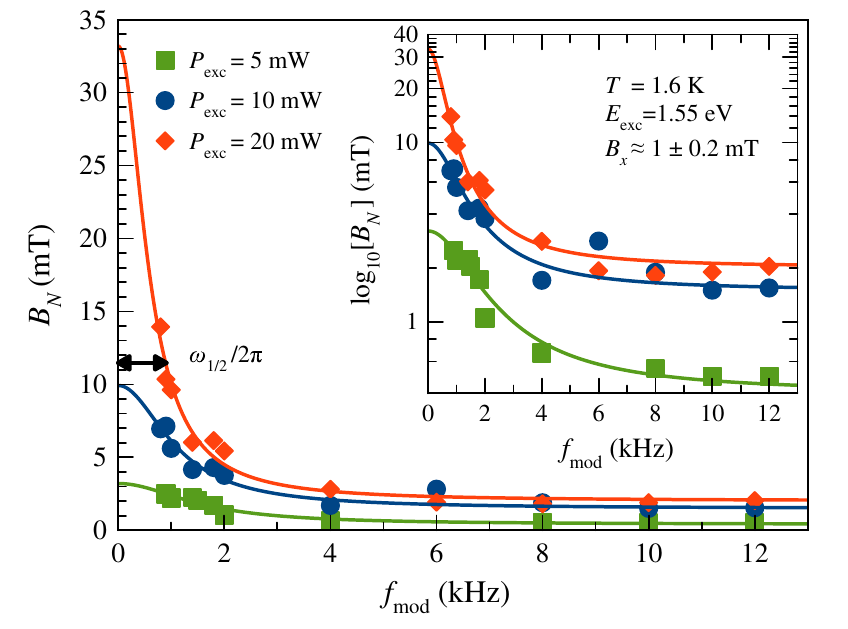}
\caption{(Color online) Calculated Overhauser fields ($B_N$) versus $f_\mathrm{mod}$. Solid lines are fits with Lorentzian functions given by Eq.~\eqref{eq_dnp_lim}, from which the values of the cut-off frequencies $\omega_{1/2}$ are extracted. The inset shows the Overhauser fields  $\log_{10}\left[ B_N \right]$ for different powers of optical excitation.
}\label{fig:Three}
\end{figure}

Next, we investigate the spin dynamics in a tilted magnetic field. 
First, scanning $B_z$, we find a tiny additional PR signal in small longitudinal fields [Fig.~\ref{fig:Two}(a)]. 
Then applying a small transverse field $B_x$ with magnitude of the order of $B_{1/2}$ and scanning  $B_z$, we find a wider PR signal [Fig.~\ref{fig:Two}(b)]. 
The results of such scans are shown in Fig.~\ref{fig:Two}(b), where a comparison of the PR curves at $B_x = 0$ and $B_x = 1$~mT is given.
Note that the narrow recovery of $\rho_c$ is present even when $B_x = 0$, which is a result of an uncompensated transverse component of the laboratory magnetic field (including the Earth field) of about $\pm 0.2$~mT.

Using the measured PR curves, we investigate the dynamics of the spin polarization recovery.
As shown in Ref.~\onlinecite{HeisterkampPRB15}, a measurement of the PR as a function of the modulation frequency, $f_\mathrm{mod}$, gives information on the dynamics of the spin system. 
As one can see from Fig.\ \ref{fig:Two}(c), the amplitude of the dip around $B_z = 0$, $A_\mathrm{PR}$, decreases with increasing $f_\mathrm{mod}$. 
We associate the observed dynamics with nuclear spin polarization: the polarization-modulated pumping results in nuclear spin cooling \cite{OO}, the cooled nuclear spins are aligned by the static transverse field and they create the Overhauser field that enhances the Hanle effect.
To evaluate the dynamics of the Overhauser field, data processing is performed following  Ref.~\onlinecite{KoturJETPLett14}.
First, a set of Hanle curves obtained at fast modulation ($f_\mathrm{mod} = 50$~kHz) is analyzed from which $\rho_0$ and $B_{1/2}$ are obtained as functions of $P_\mathrm{exc}$. 
Second, the values of $\rho_c\ (B_z = 0)$ shown in Fig.\ \ref{fig:Two}(c) are associated with the corresponding points in the Hanle dependencies, and the magnitude of the effective field $B_\mathrm{eff}$ acting on the electron spin is extracted.
We note that the PR curves shown in Fig.\ \ref{fig:Two}(c) are obtained at a fixed transverse field $B_x=1$~mT. 
Therefore, the Overhauser field can be expressed as $B_N = B_\mathrm{eff} - B_x$. 
The results of this analysis are shown in Fig.\ \ref{fig:Three}. 
As seen there, the dependencies $B_N$ versus modulation frequency $f_\mathrm{mod}$ represent decreasing functions that are expressed in the following by Eq.~\eqref{eq_dnp_norm}, and the best fit for these dependencies is provided by Lorentzian functions with characteristic widths of about $1$~kHz, which corresponds to a time on the order of a millisecond. 
The process of nuclear spin polarization is suppressed when the frequency of modulation is comparable to or larger than $T_2^{-1}$~\cite{OO}. 
The observed effect, however, develops on a somewhat longer time scale ($0.5 \lesssim \omega_{1/2}/ 2\pi \lesssim 1.5$~kHz) and is power dependent. 
In the following, we will show that the observed cutoff frequency is not $T_2^{-1}$, even though it can be related with $T_2$ and used for its experimental evaluation.

\section{\label{sec:model}Model}

The physical origin of the observed effects of modulation frequency can be understood as follows.
We point out that the observed phenomenon is measured for excitation by light with alternating helicity, creating a time-dependent non-equilibrium average spin of electrons $\braket{\mathbf S} = \mathbf{S}_0 \cos (\omega t)$, where $\mathbf{S}_0$ is the initial electron polarization and $\omega = 2 \pi f_\mathrm{mod}$.
As a result of the dynamic polarization, a time-dependent spin flow into the nuclear spin system appears given by $\mathbf{j}(t) = Q\braket{\mathbf{S}(t)}/T_{1e}$, where $Q = 4I (I+1)/3$, $I$ is the nuclear spin number and $T_{1e}$ is the time of nuclear spin relaxation by electrons via the hyperfine interaction~\cite{OO}.

The electron spin acts on the nuclear spin system also as an oscillating Knight field $\mathbf{B}_e = b_e \braket{\mathbf S}$.
Since the nuclear spins are subjected to a magnetic field, the spin flow induces an energy flow
\begin{align}
q_s (t) &= - \hbar \gamma_N \left[ \mathbf{B} + \mathbf{B}_e(t)\right] \mathbf{j}(t) \nonumber\\
&= -\frac{\hbar\gamma_N Q}{T_{1e}} \left[ B_z S_0 \cos(\omega t) + b_e S_0^2 \cos^2(\omega t)\right].
\label{eq:qs_flow}	
\end{align}
Here, $\gamma_N$ is the nuclear gyromagnetic ratio, $B_e = b_e S_0$ is the Knight field amplitude, with $b_e$ being the strength of the Knight field of a fully polarized electron, and the negative sign reflects cooling of the nuclear spin system. 
Upon averaging over the modulation period, the first term in Eq.~\eqref{eq:qs_flow} vanishes, while the second contributes to the time-averaged energy flow,
\begin{align}
\overline{q}_s =- \frac{Q}{2T_{1e}} \hbar \gamma_N b_e S^2_0. 
\label{eq:qs_aver}	
\end{align}

On the other hand, the oscillating Knight field heats up the nuclear spins.
The corresponding heating energy flow is
\begin{equation}
q_\omega (t) = -\hbar \gamma_N  \frac{d\mathbf{B}_e(t)}{dt} \mathbf{I}_B(t) = \hbar\gamma_N \omega b_e S_0 \sin(\omega t) I_B(t),
\label{eq:qw_flow}	
\end{equation}
where $I_B(t)$ is the projection of the time-dependent average nuclear spin on the direction of the Knight field. 
The time-dependent $\mathbf{I}_B(t)$ includes two contributions: $\mathbf{I}_B(t) = \mathbf{I}'_B(t) + \mathbf{I}''_B(t)$. 
The term $\mathbf{I}'_B(t)$ is induced by the Knight field via the magnetic susceptibility, $\hat{\chi}(\omega)$, of the nuclear spin system as $\mathbf{I}'_B(t) = \hat{\chi}(\omega) \mathbf{B}_e(t)$.
In turn, $\mathbf{I}''_B(t)$ results from an accumulation of the spin flow $\mathbf{j}(t)$ coming from optically pumped electrons.
The relation of $\mathbf{I}''_B(t)$ and $\mathbf{j}(t)$ is determined by relaxation of the non-equilibrium nuclear spin, and it can be written as $\mathbf{I}_B'' (t) = \int^t_0 G_N(t-\tau) \mathbf{j}(\tau) d\tau$, where the Green function, $G_N(t)$, is expressed via a correlator of the nuclear spin fluctuations~\cite{LL5} $G_N(\tau) = \frac{3}{I(I+1)} \braket{\delta I(t) \delta I(t-\tau)}$.
As follows from the definition, $G_N(0) = 1$ and, at weak fields it falls down on the time scale of $T_2$, so that $\int_0^\infty G_N(\tau)d\tau = T_2$.

Since $\mathbf{I}_B$ consists of two components, the energy flow given by Eq.~\eqref{eq:qw_flow} also has two terms: $q_\omega = q'_\omega + q''_\omega$, which contribute to the energy balance in different ways.
While (according to the fluctuation-dissipation theorem~\cite{LL5}) the imaginary part of the susceptibility in the high-temperature approximation is inversely proportional to the system temperature ($\hat{\chi}_\omega \propto \Theta_N^{-1}$), the first term, $q'_\omega (t) \propto I'_B(t)$, depends on the nuclear spin temperature $\Theta_N$ explicitly.
By averaging over the modulation period, one finds
\begin{equation}
	\overline{q}'_\omega (t) = \frac{Q}{8} (\hbar\gamma_N)^2 \omega^2 (b_e S_0)^2  \beta \hat{G}'_\omega.
	\label{eq:qw1_flow}
\end{equation}
Here, $\beta = (k_B \Theta_N)^{-1}$ and $\hat{G}'_\omega = \int_0^\infty G_N(\tau) \cos (\omega \tau)\, d\tau$.
Consequently, $\overline{q}'_\omega$ provides an additional energy relaxation channel, known as the warm-up of the nuclear spin system by the oscillating Knight field.
The second term, $q''_\omega (t) \propto I''_B(t)$, on the other hand, does not explicitly depend on $\Theta_N$: 
\begin{equation}
	\overline{q}''_\omega (t) = \frac{\hbar \gamma_N b_e Q S_0^2}{2T_{1e}} \omega \hat{G}''_\omega,
	\label{eq:qw2_flow}
\end{equation}
where $\hat{G}''_\omega = \int_0^\infty G_N(\tau) \sin (\omega \tau)\, d\tau$.
Therefore, $\overline{q}''_\omega (t)$ enters the balance equation for the inverse spin temperature $\beta$ as a source, similarly to $\overline{q}_s$.
The balance equation reads 
\begin{equation}
\frac{\partial\beta}{\partial t} = - \frac{\beta}{T_1} + \frac{\overline{q}_s + \overline{q}'_\omega + \overline{q}''_\omega}{C_N},
\label{eq_balance}	
\end{equation}
where $C_N = \frac{1}{3}I(I+1) (\hbar\gamma_N)^2 (B^2+B_L^2)$ is the heat capacity of the nuclear spin system.
Taking $\partial\beta/\partial t = 0$ we obtain the following expression for the inverse spin temperature
\begin{widetext}
\begin{equation}
\beta = - \left( 1 + \frac{1}{2} \frac{b_e^2 S_0^2}{B^2+B_L^2} \omega T_1 \hat{G}_\omega'\right)^{-1} \frac{3T_1b_eQS_0^2}{2I(I+1) T_{1e} \hbar \gamma_N (B^2+B_L^2)} \Bigl(1-\omega \hat{G}_\omega''\Bigr).
\label{eq_beta}	
\end{equation}
The steady-state nuclear spin polarization corresponding to $\beta$ and established in the external magnetic field is given by
\begin{equation}
\frac{\braket{I}}{I} = \frac{I+1}{3}\hbar\gamma_N \beta B =
- \frac{B b_e Q S_0^2 T_1}{2I T_{1e}}	
\left( 1 + \frac{1}{2}\frac{b_e^2 S_0^2}{B^2 + B_L^2 } \omega^2 T_1 \hat{G}_\omega' \right)^{-1}
\frac{1 - \omega \hat{G}_\omega''}{B^2 + B_L^2}.
\label{eq_dnp}
\end{equation}
\end{widetext}

In the vicinity of a donor, $T_1 \approx T_{1e}$, and we estimate its value according to Ref.~\onlinecite{DPJetp75},
\begin{equation}
\frac{1}{T_{1e}} = \frac{2}{3} S(S+1) b_e^2 \gamma_N^2 \tau_c \frac{B^2+\xi B_L^2}{B^2 + B_L^2},
\label{eq_T1e}
\end{equation}
where $\xi \leq 3$.
Remarkably, the steady-state nuclear spin polarization scales linearly with $b_e$, as follows from Eq.~\eqref{eq_dnp} when $T_1$ is determined by Eq.~\eqref{eq_T1e}. Thus, the normalized frequency dependence
\begin{equation}
\frac{\braket{I}_\omega}{\braket{I}_0} = \left( 1 + \omega^2 \hat{G}_\omega' \frac{S_0^2}{\gamma_N^2(B^2 + \xi B_L^2)\tau_c} \right)^{-1} \Bigl(1 - \omega \hat{G}_\omega''\Bigr)
\label{eq_dnp_norm}
\end{equation}
does not contain $b_e$, and, thus it is not sensitive to the shape of the wave function of the donor-bound electron. 
Hence, Eq.~\eqref{eq_dnp_norm} can be used universally.

At high frequencies of modulation, $\lim_{\omega\to\infty} \omega \hat{G}_\omega'' = G_N (0) = 1$.
Therefore, $\braket{I}_\omega$ tends to zero. 
For this reason, no cooling of the nuclear spin ensemble is possible at high modulation frequency, i.\,e., at $\omega \gg T_2^{-1}$. 
The exception is the case when a strong transverse field is applied and $G_N (t)$ oscillates at the frequency of the nuclear magnetic resonance.
In this case, resonant cooling is observed~\cite{OO}.

\begin{figure}[t]
\includegraphics[clip, width=\columnwidth]{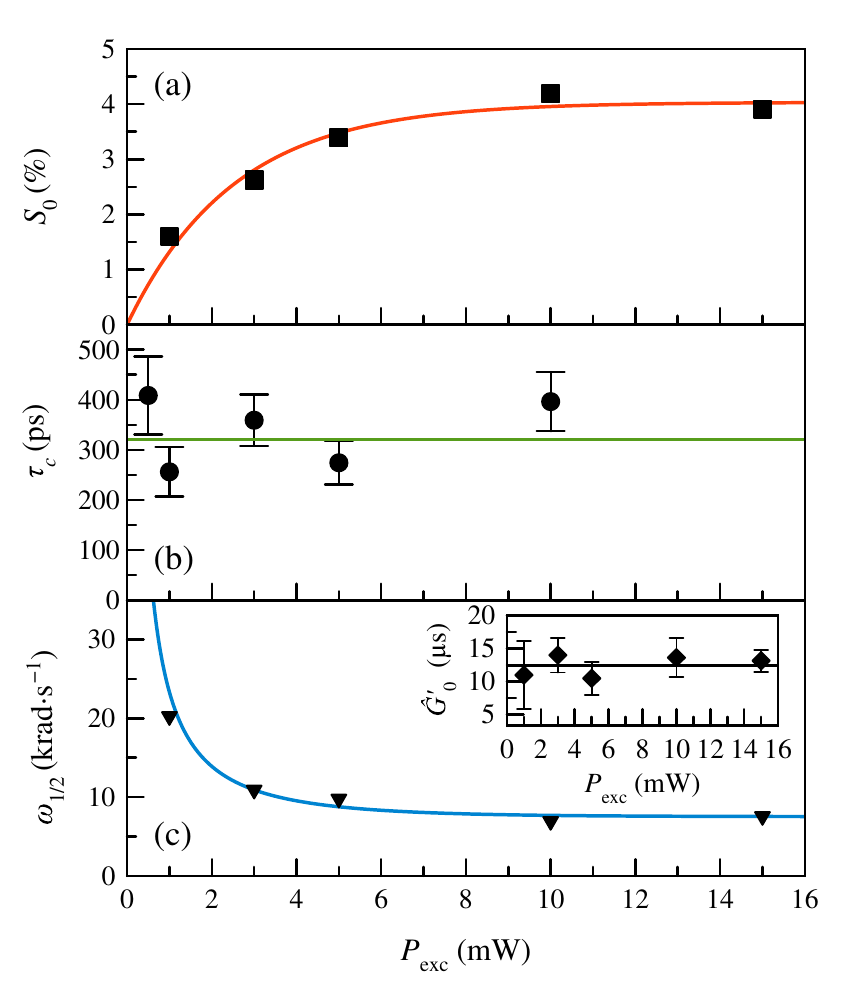}
\caption{(Color online) (a) Electron mean spin polarization (squares) and its fit with a saturating exponential function: $S_0 (P_\mathrm{exc}) = S_\infty [1 - \exp(-P_\mathrm{exc}/P_0)]$ (solid line). The fit parameters are: $S_\infty = 0.041$ and $P_0 = 2.6$~mW. (b) Power dependence of the electron correlation time. The solid green line shows a linear fit with~$\tau_c = 320$~ps. (c)~Cut-off frequency of the Overhauser field build-up versus excitation power $P_\mathrm{exc}$ (triangles). The solid line shows a fit of the data with Eq.~\eqref{eq_w12}. Inset shows the corresponding nuclear spin correlator $\hat{G}'_0$ as a result of normalization of $\omega_{1/2}$ to $S_0$. The solid black line shows a linear fit with $\hat{G}'_0 = 12$~\textmu{}s.
}\label{fig:Four}
\end{figure}

At low frequencies of modulation, $\omega \ll T_2^{-1}$, $1 - \omega \hat{G}_\omega'' \simeq 1$ and $\hat{G}_\omega' \simeq \hat{G}'_0$.
As a result, Eq.~\eqref{eq_dnp_norm} simplifies to
\begin{equation}
\frac{\braket{I}_\omega}{\braket{I}_0} = \frac{1}{1 + \omega^2/\omega_{1/2}^2},
\label{eq_dnp_lim}	
\end{equation}
where the cut-off frequency is given by
\begin{equation}
\omega_{1/2} = \frac{1}{S_0} \sqrt{\frac{\tau_c}{\hat{G}'_0}} \gamma_N \sqrt{B^2 + \xi B_L^2}.
\label{eq_w12}
\end{equation}
Note that $\omega_{1/2}^{-1}$ is not equal to any nuclear spin relaxation time, neither $T_1$ nor $T_2$, but it is related to $T_2$ via the zero-frequency Fourier component of $G_N$. 

The magnitude of $S_0$ can be evaluated from the experiment provided that $\rho_0 = S_0$.
As shown in Fig.~\ref{fig:Four}(a), the power dependence of $S_0$ demonstrates saturation that can be described by a single exponent, as follows from a simple rate equation for the populations of the spin-up and spin-down states subject to a generation term $P_\mathrm{exc}$.
The frequency dependencies of the Overhauser field shown in Fig.~\ref{fig:Three} are fitted with Lorentzian functions, from which the cut-off frequencies are extracted as function of the excitation power $P_\mathrm{exc}$.
Since $\tau_c$ does not depend on $P_\mathrm{exc}$, as shown in Fig.~\ref{fig:Four}(b), the power dependence of $\omega_{1/2}$ is only mediated by $S_0$ and represents a curve decaying with increasing $P_\mathrm{exc}$, as shown by the triangles in Fig.~\ref{fig:Four}(c). 
Equation~\eqref{eq_w12} can be used to fit this dependence with a single variable parameter $\hat{G}'_0\approx12$~\textmu{}s, as displayed by the solid line in this figure. 
In this fitting, we take $\gamma_N/(2\pi) \approx 9.3$ kHz/mT as an estimate for the nuclear gyromagnetic ratio averaged over all nuclear species ($^{75}$As, $^{69}$Ga, $^{71}$Ga with weights $0.5$, $0.3$ and $0.2$, respectively~\cite{HarrisDEG09}), $\xi = 3$, and $B_L = 0.15$ mT~\cite{VladimirovaPRB17}. 

The fit of the experimental data allows us to determine $\hat{G}'_0$. 
Note that when $B \ll B_L$, $\hat{G}'_0 \approx T_2$. 
At $B = 0$, the correlator $G'_\omega$ is centered at zero frequency, and at field $B>B_L$ its maximum is shifted to the Larmor frequency in field $B$. 
Therefore, the zero-frequency value, in this condition, does not determine $T_2$, and the relaxation time should be extracted from the spectrum $G'_\omega$. 
To evaluate $T_2$ from our experimental data we use a simple model of the nuclear spin correlator. 
Consider the correlator in a weak external magnetic field to have the form of a decaying oscillation $G_N(t) = \cos(\Omega_L t) \exp(-t/T_2)$, with the Fourier components being:
\begin{subequations}
\begin{align}
\hat{G}'_\omega &= \frac{T_2}{2} \left( \frac{1}{1+(\Omega_L+\omega)^2T_2^2} + \frac{1}{1+(\Omega_L-\omega)^2T_2^2} \right),\\
\hat{G}''_\omega &= \frac{T_2}{2} \left( \frac{T_2(\Omega_L+\omega)}{1+(\Omega_L+\omega)^2T_2^2} + \frac{T_2(\Omega_L-\omega)}{1+(\Omega_L-\omega)^2T_2^2} \right),
\end{align}	
\end{subequations}
where $\Omega_L = \gamma_N B$ is the nuclear Larmor frequency.

\begin{figure}[t]
\includegraphics[clip, width=\columnwidth]{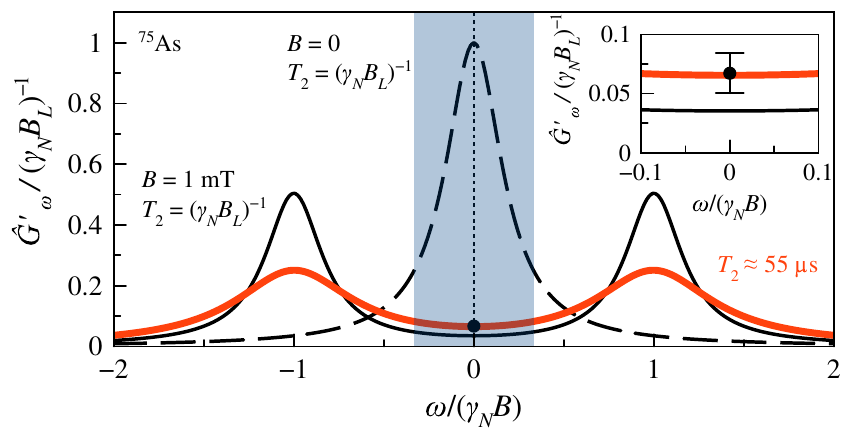}
\caption{(Color online) Normalized real Fourier component of the correlator of the nuclear-nuclear interactions versus frequency for $B = 0$ (dashed line) and $B = 1$~mT (solid lines) calculated for the $^{75}\mathrm{As}$ nuclear species. Filled area defines the range of  frequencies $\omega < \omega_{1/2}$ observed in the experiment. The black filled circle represents the value of the correlator determined from experiment [see inset on the Fig.~\ref{fig:Four}(c)]. The thick red line represents the $\hat{G}'_\omega$ calculated for the best fit of $G'_0$ ($T_2 = 55$~\textmu{}s) and the thin, solid and dashed black lines are calculated for the spin-spin relaxation time $T_2 = (\gamma_N B_L)^{-1} \approx114$~\textmu{}s.
}\label{fig:Five}	
\end{figure}

Within this approximation, the fitting parameter $\hat{G}'_0$ found from the experimental data makes it possible to evaluate the nuclear spin relaxation time $T_2$ in a weak magnetic field. 
Since the correlator already at $B_x=1$~mT demonstrates clearly a resonant behavior (Fig.~\ref{fig:Five}), its value near zero frequency is determined by the isotope with the smallest gyromagnetic ratio: $\gamma_N[^{75}\text{As}]=4.596\times10^7$~rad/(Ts), and $\gamma_N[^{69}\text{Ga}]=6.439\times10^7$~rad/(Ts), $\gamma_N[^{71}\text{Ga}]=8.181\times10^7$~rad/(Ts), i.\,e. arsenic. 
One can plot the frequency dependence $\hat{G}'_\omega$ calculated for $B = 0$ and 1~mT (Fig.~\ref{fig:Five}). 
Thereby, we calculate $T_2 = (\gamma_N B_L)^{-1}\approx 114$~\textmu{}s, which agrees with the commonly accepted values $T_2 \sim 10^{-4}$~s\ \cite{PagetPRB77}.
As one can see from the figure, the correlator has the maximum value $\hat{G}'_0 = T_2$ when the external field is zero.
If the field is applied, the Fourier maxima are shifted to the positive and negative frequency values of the Larmor precession, and $\hat{G}'_\omega$ drops.
Its value extracted from the experiment is, however, still larger than the simple model predicts (point in Fig.~\ref{fig:Five}).
Therefore, an extended model for the correlator $G_N(t)$ needs to be developed. 
Since this is far beyond the scope of the current work, we simply evaluate the value of $T_2$, which explains the experimental value of $G'_0 = 12$~\textmu{}s.
Note that for the frequencies of modulation used in all our experiments (see filled area in Fig.~\ref{fig:Five}), the correlator $G'_\omega$ changes weakly, and its value is approximately equal to $G'_0$ within the fitting inaccuracy. This justifies \emph{a~posteriori} the assumptions of our theoretical model. 
We find this value to be $T_2 = 55$~\textmu{}s, which is two times smaller than the one estimated through $B_L\approx0.15$~mT \citep{PagetPRB77} that was used to evaluate the complex behavior of the nuclear spin relaxation in NMR~\citep{ChenPRL11, OnoPRB14}. 
It is also several times smaller than the $T_2=100$~\textmu{}s value measured in GaAs/AlGaAs quantum wells~\cite{SanadaPRL06} and $T_2=270$~\textmu{}s for lattice matched GaAs/AlGaAs quantum dots~\cite{MakhoninNmat11}, where the measurements were done at $B \sim 1$~T.
The most likely origin of this difference is that in our weak-field experiments, the external magnetic field $B$ is comparable to the local field $B_L$, and, therefore, the non-secular part of the dipole-dipole interaction may come into play, thus increasing the rate of the nuclear spin relaxation.

\section{\label{sec:conclusion}Conclusion}
In summary, the spin relaxation of the nuclear spin ensemble has been studied in $n$-doped GaAs crystal using a modified spin inertia method.
We find that optical pumping with light of alternating helicity induces a fast build-up of the Overhauser field.
The dynamics is observed on a sub-second time scale showing a frequency cut-off that varies by several times upon increasing the pumping power.
The experimental results are interpreted within a developed model, which predicts a drop of the nuclear spin polarization when the light helicity modulation rate reaches a characteristic frequency $\omega_{1/2}$, determined by the spin correlation time of donor-bound electrons and the nuclear spin-spin relaxation time $T_2$, which was estimated as $T_2 = 55$~\textmu{}s, i.e. noticeably shorter than the $T_2 \sim 2 \times 10^{-4}$~s determined by NMR methods at high magnetic fields.

\begin{acknowledgements}
We acknowledge the financial support of the Deutsche Forschungsgemeinschaft in the frame of the ICRC TRR 160 (Project No. A6) and the Russian Foundation for Basic Research (Project No. 15-52-12020).
\end{acknowledgements}


\begin{thebibliography}{}
\bibitem{HennebergerQBits} \emph{Semiconductor Quantum Bits}, edited by F.\ Henneberger and O.\ Benson \href{\doibase10.4032/9789814241199}{(Pan Stanford, Singapore, 2008)}. 

\bibitem{DyakonovSpinBook} \emph{Spin Physics in Semiconductors}, 2nd Ed., edited by M.~I.\ Dyakonov  \href{\doibase10.1007/978-3-319-65436-2}{(Springer, Cham, 2017)}. 

\bibitem{UrbaszekRMP13} B.\ Urbaszek, X.\ Marie, T.\ Amand, O.\ Krebs, P.\ Voisin, P.\ Maletinsky, A.\ H\"ogele, and A.\ Imamoglu, ``Nuclear spin physics in quantum dots: An optical investigation'', \href{\doibase10.1103/RevModPhys.85.79}{Rev.\ Mod.\ Phys.\ \textbf{85}, 79 (2013)}.

\bibitem{GoldmannSpinTemp} M.\ Goldman, \emph{Spin Temperature and Nuclear Magnetic Resonance in Solids} (Clarendon Press, Oxford, 1970).

\bibitem{OO} I.~A.\ Merkulov and V.~G.\ Fleisher, ``Optical orientation of coupled electron-nuclear spin system of a semiconductor'', in \emph{Optical Orientation}, edited by F.\ Meier and B.~P.\ Zakharchenya (North-Holland, Amsterdam, 1984), Chap.\ 5.

\bibitem{PagetPRB77} D.\ Paget, G.\ Lampel, B.\ Sapoval, and V.~I.\ Safarov, ``Low field electron-nuclear spin coupling in gallium arsenide under optical pumping conditions'', \href{\doibase10.1103/PhysRevB.15.5780}{Phys.\ Rev.\ B\ \textbf{15}, 5780 (1977)}.

\bibitem{MerkulovFTT98} I.~A.\ Merkulov, ``Formation of a nuclear spin polaron under optical orientation in GaAs-type semiconductors'', Fiz.\ Tverd.\ Tela\ (St.~Petersburg) \textbf{40}, 1018 (1998) [\href{\doibase10.1134/1.1130450}{Phys.\ Solid\ State\ \textbf{40}, 930 (1998)}].

\bibitem{KalevichIzv82} V.~K. Kalevich, V.~D.\ Kulkov, and V.~G. Fleisher, ``Optical cooling of the nuclear spin system of a semiconductor in a conjunction with the adiabatic demagnetization'', Izv.\ Akad.\ Nauk\ SSSR\ Ser.\ Fiz.\ \textbf{46}, 492 (1982); [Bull.\ Acad.\ Sci.\ USSR,\ Phys.\ Ser.\ \textbf{46}, 70 (1982)].

\bibitem{VladimirovaArXiv17} M.\ Vladimirova, S.\ Cronenberger, D.\ Scalbert, I.~I.\ Ryzhov, V.~S.\ Zapasskii, G.~G.\ Kozlov, A.\ Lema\^itre, and K.~V.\ Kavokin, ``Spin temperature concept verified by optical magnetometry of nuclear spins'', \href{http://arxiv.org/abs/1706.02528}{arXiv:1706.02528}.

\bibitem{PagetPRB08} D.\ Paget, T.\ Amand, and J.~P.\ Korb, ``Light-induced nuclear quadrupolar relaxation in semiconductors'', \href{\doibase10.1103/PhysRevB.77.245201}{Phys.\ Rev.\ B\ \textbf{77}, 245201 (2008)}.

\bibitem{KoturPRB16} M.\ Kotur, R.~I. Dzhioev, M.\ Vladimirova, B.\ Jouault, V.~L.\ Korenev, and K.~V.\ Kavokin, ``Nuclear spin warm up in bulk \emph{n}-GaAs'', \href{\doibase10.1103/PhysRevB.94.081201}{Phys.\ Rev.\ B\ \textbf{94}, 081201(R) (2016)}.

\bibitem{EickhoffPRB02} M.\ Eickhoff, B.\ Lenzman, G.\ Flinn, and D.\ Suter, ``Coupling mechanisms for optically induced NMR in GaAs quantum wells'', \href{\doibase10.1103/PhysRevB.65.125301}{Phys.\ Rev.\ B\ \textbf{65}, 125301 (2002)}.

\bibitem{GiriPRL13} R.\ Giri, S.\ Cronenberger, M.~M.\ Glazov, K.~V.\ Kavokin, A.\ Lema\^itre, J.\ Bloch, M.\ Vladimirova, and D.\ Scalbert, ``Nondestructive measurement of nuclear magnetization by off-resonant Faraday rotation'', \href{\doibase10.1103/PhysRevLett.111.087603}{Phys.\ Rev.\ Lett.\ \textbf{111}, 087603 (2013)}.

\bibitem{MalinowskiPRL17} F.~K.\ Malinowski, F.\ Martins, \L. Cywi\'nski, M.~S.\ Rudner, P.~D.\ Nissen, S. Fallahi, G.~C.\ Gardner, M.~J.\ Manfra, C.~M.\ Marcus, and F. Kuemmeth, ``Spectrum of the nuclear environment for GaAs spin qubits'', \href{\doibase10.1103/PhysRevLett.118.177702}{Phys.\ Rev.\ Lett.\ \textbf{118}, 177702 (2017)}.

\bibitem{KikkawaSci00} J.~M.\ Kikkawa and D.~D.\ Awschalom, ``All-optical magnetic resonance in semiconductors'', \href{\doibase10.1126/science.287.5452.473}{Science\ \textbf{287}, 483 (2000)}.

\bibitem{GreilichSci06} A.\ Greilich, D.~R.\ Yakovlev, A.\ Shabaev, Al.~L.\ Efros, I.~A.\ Yugova, R.\ Oulton, V.\ Stavarache, D.\ Reuter, A.\ Wieck, and M.\ Bayer, ``Mode locking of electron spin coherence in singly charged quantum dots'', \href{\doibase10.1126/science.1128215}{Science\ \textbf{313}, 341 (2006)}.

\bibitem{GreilichSci07} A.\ Greilich, A.\ Shabaev, D.~R.\ Yakovlev, Al.~L.\ Efros, I.~A.\ Yugova, D.\ Reuter, A.~D.\ Wieck, and M.\ Bayer, ``Nuclei-induced focusing of electron spin coherence'', \href{\doibase10.1126/science.1146850}{Science\ \textbf{317}, 1896 (2007)}.

\bibitem{FalkPRL15} A.~L.\ Falk, P.~V.\ Klimov, V.\ Iv\'ady, K.\ Sz\'asz, D.~J.\ Christle, W.~F.\ Koehl, \'A.\ Gali, and D.~D.\ Awschalom, ``Optical polarization of nuclear spins in Silicon Carbide'', \href{\doibase10.1103/PhysRevLett.114.247603}{Phys.\ Rev.\ Lett.\ \textbf{114}, 247603 (2015)}.

\bibitem{ChekhovichNmat17} E.~A.\ Chekhovich, A.\ Ulhaq, E.\ Zallo, D.\ Ding, O.~G.\ Schmidt, and M.~S.\ Skolnick, ``Measurement of the spin temperature of optically cooled nuclei and GaAs hyperfine constants in GaAs/AlGaAs quantum dots'', \href{\doibase10.1038/nmat4959}{Nat.\ Mater.\ \textbf{16}, 982 (2017)}.

\bibitem{HeisterkampPRB15} F.\ Heisterkamp, E.~A.\ Zhukov, A.\ Greilich, D.~R.\ Yakovlev, V.~L.\ Korenev, A.\ Pawlis, and M.\ Bayer, ``Longitudinal and transverse spin dynamics of donor-bound electrons in fluorine-doped ZnSe: Spin inertia versus Hanle effect'', \href{\doibase10.1103/PhysRevB.91.235432}{Phys.\ Rev.\ B\ \textbf{91}, 235432 (2015)}.

\bibitem{DzhioevPRB02} R.~I.\ Dzhioev, K.~V.\ Kavokin, V.~L.\ Korenev, M.~V.\ Lazarev, B.~Ya.\ Meltser, M.~N.\ Stepanova, B.~P.\ Zakharchenya, D.\ Gammon, and D.~S.\ Katzer, ``Low-temperature spin relaxation in \emph{n}-type GaAs'', \href{\doibase10.1103/PhysRevB.66.245204}{Phys.\ Rev.\ B\ \textbf{66}, 245204 (2002)}.

\bibitem{DzhioevPRL02} R.~I.\ Dzhioev, V.~L.\ Korenev, I.~A.\ Merkulov, B.~P.\ Zakharchenya, D.\ Gammon, Al.~L.\ Efros, and D.~S.\ Katzer, ``Manipulation of the spin memory of electrons in \emph{n}-GaAs'', \href{\doibase10.1103/PhysRevLett.88.256801}{Phys.\ Rev.\ Lett.\ \textbf{88}, 256801 (2002)}.

\bibitem{WeisbuchPRB77} C.\ Weisbuch  and C.\ Hermann, ``Optical detection of conduction-electron spin resonance in  GaAs, Ga$_{1-x}$In$_x$As, and Ga$_{1-x}$Al$_x$As'', \href{\doibase10.1103/PhysRevB.15.816}{Phys.\ Rev.\ B\ \textbf{15}, 816 (1977)}.

\bibitem{KoturJETPLett14} M.\ Kotur, R.~I.\ Dzhioev, K.~V.\ Kavokin, V.~L.\ Korenev, B.~R.\ Namozov, P.~E.\ Pak, and Yu.~G.\ Kusrayev, ``Nuclear spin relaxation mediated by Fermi-edge electrons in \emph{n}-type GaAs'', \href{\doibase10.1134/S0021364014010068}{JETP\ Lett.\ \textbf{99}, 40 (2014)}.

\bibitem{LL5} L.~D.\ Landau and E.~M.\ Lifshitz, \emph{Statistical Physics. Part~I}, Course of Theoretical Physics, 3rd\ Ed., Vol.\ 5, Ch.~XII (Elsevier, Oxford 1980).

\bibitem{DPJetp75} M.~I.\ Dyakonov and V.~I.\ Perel, ``Cooling of a system of nuclear spins following optical orientation of electrons in semiconductors'', Zh.\ Eksp.\ Teor.\ Fiz.\ \textbf{68}, 1514 (1975) [\href{http://www.jetp.ac.ru/cgi-bin/e/index/e/41/4/p759?a=list}{Sov.\ Phys.\ JETP\ \textbf{41}, 759 (1975)}].

\bibitem{HarrisDEG09} R.~K.\ Harris, E.~D.\ Becker, S.~M.\ Cabral de Menezes, R.\ Goodfellow, and P.\ Granger ``NMR nomenclature. Nuclear spin properties and conventions for chemical shifts'', \href{\doibase10.1002/mrc.1042}{Pure\ Appl.\ Chem.\ \textbf{73}, 1795 (2001)}.

\bibitem{VladimirovaPRB17} M.\ Vladimirova, S.\ Cronenberger, D.\ Scalbert, M.\ Kotur, R.~I.\ Dzhioev, I.~I.\ Ryzhov, G.~G.\ Kozlov, V.~S.\ Zapasskii, A.\ Lema\^itre, and K.~V.\ Kavokin, ``Nuclear spin relaxation in \emph{n}-GaAs: From insulating to metallic regime'', \href{\doibase10.1103/PhysRevB.95.125312}{Phys.\ Rev.\ B\ \textbf{95}, 125312 (2017)}.

\bibitem{ChenPRL11} Y.~S.\ Chen, D.\ Reuter, A.~D.\ Wieck, and G.\ Bacher, ``Dynamic nuclear spin resonance in \emph{n}-GaAs'', \href{\doibase10.1103/PhysRevLett.107.167601}{Phys.\ Rev.\ Lett.\ \textbf{107}, 167601 (2011)}.

\bibitem{OnoPRB14} M.\ Ono, J.\ Ishihara, G.\ Sato, S.\ Matsuzaka, Y.\ Ohno, and H.\ Ohno, ``Strain and origin of inhomogeneous broadening probed by optically detected nuclear magnetic resonance in a (110) GaAs quantum well'', \href{\doibase10.1103/PhysRevB.89.115308}{Phys.\ Rev.\ B\ \textbf{89}, 115308 (2014)}.

\bibitem{SanadaPRL06} H.\ Sanada, Y.\ Kondo, S.\ Matsuzaka, K.\ Morita, C.~Y.\ Hu, Y.\ Ohno, and H.\ Ohno, ``Optical pump-probe measurements of local nuclear spin coherence in semiconductor quantum wells'', \href{\doibase10.1103/PhysRevLett.96.067602}{Phys.\ Rev.\ Lett.\ \textbf{96}, 067602 (2006)}.

\bibitem{MakhoninNmat11} M.~N.\ Makhonin, K.~V.\ Kavokin, P.\ Senellart, A.\ Lema\^itre, A.~J.\ Ramsay, M.~S.\ Skolnick, and A.~I.\ Tartakovskii, ``Fast control of nuclear spin polarization in an optically pumped single quantum dot'', \href{\doibase10.1038/nmat3102}{Nat.\ Mater.\ \textbf{10}, 844 (2011)}.

\end{thebibliography}
\end{document}